\documentclass[12pt]{article}
\usepackage{amssymb}
\usepackage{amsmath}
\usepackage{latexsym}
\usepackage{array}
\usepackage{morefloats}
\usepackage{epsfig}
\usepackage{rotating}
\usepackage{graphicx}
\usepackage{subfigure}
\usepackage{url}
\usepackage{enumerate}
\usepackage{wasysym}
\usepackage{threeparttable}
\usepackage{lscape}
\usepackage{natbib}
\usepackage{color}
\usepackage{epstopdf}
\usepackage{hyperref}
\newcommand{\mbv}[1]{\mbox{\boldmath$#1$\unboldmath}}
\newcommand{\mbf}[1]{\mathbf{#1}}

\newcommand{\mcal}[1]{\mathcal{#1}}

\newcommand{\qed}{\nobreak \ifvmode \relax \else
      \ifdim\lastskip<1.5em \hskip-\lastskip
      \hskip1.5em plus0em minus0.5em \fi \nobreak
      \vrule height0.75em width0.5em depth0.25em\fi}

\AppendGraphicsExtensions{.tif}

\def\log{\hbox{log}}

\def\boxit#1{\vbox{\hrule\hbox{\vrule\kern6pt
          \vbox{\kern6pt#1\kern6pt}\kern6pt\vrule}\hrule}}

\def\be{\begin{eqnarray}}
\def\ee{\end{eqnarray}}
\def\bq{\begin{equation}}
\def\eq{\end{equation}}
\def\bse{\begin{eqnarray*}}
\def\ese{\end{eqnarray*}}

\begin{document}
\thispagestyle{empty} \baselineskip=28pt

\begin{center}
{\LARGE{\bf A Bayesian Semiparametric Jolly-Seber Model with Individual Heterogeneity: An Application to Migratory Mallards at Stopover}}
\end{center}

\baselineskip=12pt
\vskip 2mm
\begin{center}
Guohui Wu\footnote{\baselineskip=10pt
(\baselineskip=10pt to whom correspondence should be addressed) SAS Institue Inc., 100 SAS Campus Drive, Cary, NC 27513, raywu2014@gmail.com},
Scott H. Holan\footnote{\baselineskip=10pt Department of Statistics, University of Missouri,146 Middlebush Hall, Columbia, MO 65211-6100},
Alexis Avril\footnote{\baselineskip=10pt Centre for Ecology and Evolution in Microbial Model Systems, Linnaeus University, SE-391 82 Kalmar, Sweden},
and Jonas Waldenstr\"{o}m$^3$
\end{center}

%%
%%
%%%%%%%%%%%%%%%%%%%%%%%%%%%%%%%%%%%%%%%%%%%%%%%%%%%%%%%%%%%%%%%%%%%%%%%%
\vskip 2mm
\begin{center}
{\large{\bf Abstract}}
\end{center}
        We propose a Bayesian hierarchical Jolly-Seber model that can account for individual heterogeneity in departure and the dependence of arrival time on covariates. Additionally, our model provides a semiparametric functional form for modeling capture probabilities. The model is flexible and can be used to estimate the stopover duration and stopover population size, which are key to stopover duration analysis. From the modeling perspective, our model allows for individual heterogeneity in departure due to a continuous intrinsic factor that varies with time and individual. A stochastic process is considered to model the change of this intrinsic factor over time. Moreover, our model links extrinsic factors to capture probabilities and arrival time. Consequently, our proposed model enables us to draw inference about the impacts of the intrinsic factor on departure, and extrinsic factors on both capture outcome and arrival time. Through the use of a semiparametric model for capture probabilities, we allow the data to suggest the functional relationship between extrinsic factors and capture probabilities rather than relying on an imposed parametric model. By using data augmentation, we develop a well customized Markov chain Monte Carlo algorithm that is free of tuning. We demonstrate the effectiveness of our model through a motivating example of stopover duration analysis for mallards (\textit{Anas platyrhynchos}) studied during fall migration in Sweden. 
\baselineskip=12pt

%%%%%%%%%%%%%%%%%%%%%%%%%%%%%%%%%%%%%%%%%%%%%%%%%%%%%%%%%%%%%%%%%%%%%%%%
%
%
%

\baselineskip=12pt
\par\vfill\noindent
\par\medskip\noindent
\clearpage\pagebreak\newpage \pagenumbering{arabic}
\baselineskip=24pt
%\singlespacing
\section{Introduction}
      Migration is a common phenomenon in birds, especially in areas with pronounced seasonal variation. However, in most species, migration is not conducted as a single flight from the breeding area to the non-breading area; rather it is broken down into shorter legs interspersed with stopovers of variable length at suitable sites where energy spent during migration can be replenished \citep[e.g., see][and the references therein]{newton2010migration}. Mostly determined by the time spent at stopover sites, the overall speed of migration is tightly linked to behaviors at stopover sites, and the distribution and quality of stopover sites impacts the success and survival of birds during migration. A key to stopover duration analysis rests on understanding various species-specific stopover behaviors and how intrinsic and extrinsic factors contribute to these behaviors. For this reason, primary objectives in stopover studies are to estimate the timing of arrival and departure, stopover duration (i.e., the length of stay at a stopover site), stopover population size, and to understand the impacts of intrinsic and extrinsic factors. To accomplish these objectives, capture-recapture studies have been used extensively over the past few decades, with a variety of models being proposed for stopover duration analysis \citep[e.g., see][and the references therein]{pledger2009stopover,king2010bayesian,matechou2010applications}.

      The two most commonly used capture-recapture models for stopover duration analysis are the Cormack-Jolly-Seber (CJS) \citep{cormack1964estimates} and Jolly-Seber (JS) models \citep{jolly1965explicit,seber1965note}. Among the many underlying assumptions for the CJS and JS models, two important assumptions are: (1) every individual that is captured needs to be correctly and uniquely marked; (2) every individual that is alive and present in the study area has an equal likelihood of capture and survival (i.e., homogeneous capture probabilities and survival probabilities) \citep{williams2002analysis}. The fundamental difference between the CJS and JS models is that the former conditions on the first capture while the latter does not. In relation to stopover duration analyses, the CJS model allows estimation of survival probabilities (i.e., stopover retention probabilities in stopover duration analysis), based on which one can adopt the life expectancy equation \citep{seber1982} to derive an estimator of the stopover duration \citep[e.g., see][]{morris2006utility}. To exemplify this, \citet{kaiser1995estimating}, \citet{dinsmore2003influence}, and \citet{rice2007local} demonstrate applications of the CJS model for estimating stopover duration. Importantly, the resulting estimate of the stopover duration from the CJS model can be biased due to the conditional nature of the model and unknown arrival time \citep{pledger2009stopover}.

      Unlike its CJS counterpart, in addition to estimating capture probabilities and survival probabilities, the JS model can be used to estimate population size and entrance probabilities (i.e., the probability of entering the study area right before each sampling period). \cite{schwarz1996general} present a general, yet flexible, formulation of the JS model that is advantageous in the sense that their approach explicitly incorporates the entrance probabilities into the likelihood function. As a result, the Schwarz and Arnason (SA) formulation of the JS model allows for a versatile modeling framework capable of imposing restrictions or incorporating covariates for the entrance probabilities. Moreover, it is shown that unbiased estimators for the entrance probabilities and their derived quantities can be achieved in the presence of heterogeneous capture probabilities \citep[see][and the references therein]{schwarz2001jolly}. Based on the SA formulation, \cite{royle2008hierarchical} provide a state-space formulation of the JS model under the Bayesian hierarchical modeling paradigm. In this setup, data augmentation \citep{tanner1987calculation} is considered to facilitate Bayesian model estimation using freely available software such as WinBUGS \citep{lunn2000winbugs}.

      Building upon the SA formulation of the JS model, \citet{pledger2009stopover} develop a flexible stopover model under the frequentist framework to allow capture and stopover retention probabilities to depend on an unknown time since arrival . Apart from deriving indirect estimate of the mean stopover duration, they also consider modeling the stopover retention curve to examine different stopover departure patterns. To extend the stopover model by \citet{pledger2009stopover}, \citet{matechou2014monitoring} develop a mixture model where captured individuals do not need to be correctly and distinctly marked. In other words, data for such an extended model consists of counts of individuals captured in each sampling period rather than encounter histories of uniquely marked individuals. Subsequently, \citet{lyons2015population} develop a Bayesian stopover model that accommodates both encounter histories of uniquely marked individuals and counts of unmarked individuals. Their model allows for the estimation of capture and stopover retention probabilities, entrance probabilities, stopover population size, and stopover duration. In particular, the estimator of the stopover duration is derived from latent state variables that are introduced via data augmentation, following \citet{royle2008hierarchical}. Recently, \citet{matechou2016bayesian} develop a stopover model by extending the JS model to allow individuals to arrive in different groups and hence their model accounts for heterogeneity in departure due to a group effect. Additionally, to address individual heterogeneity in arrival time due to a group effect, entrance probabilities are modeled using a finite mixture.

      Despite the usefulness of the aforementioned stopover models, many real-world applications require development of a data-specific model. As in our motivating example, there is a need to address individual heterogeneity in migratory bird departure decisions due to a continuous intrinsic factor that varies with both time and individual. In addition, there is also a need to link the arrival time and capture probabilities to extrinsic factors and to infer the functional relationship between them. As a consequence, we develop a stopover model using data augmentation under the Bayesian hierarchical state-space framework. The methodological contributions can be described as follows. First, our model accounts for individual heterogeneity in departure due to a time-varying continuous individual covariate. Second, our model allows for a data-driven functional relationship between the capture probabilities and extrinsic factors through the use of smoothing splines, which enables us to detect a nonlinear temporal trend. Furthermore, our model links the arrival time to extrinsic factors and hence allows us to draw inference about their impacts on the time of arrival. More importantly, we develop a well-tailored Markov chain Monte Carlo (MCMC) algorithm for our proposed model to avoid tedious user-defined tuning.

      This paper is organized as follows. Section~\ref{sec:data} introduces the motivating data from mallard monitoring study. Section~\ref{sec:methods} presents the proposed state-space model and provides two goodness-of-fit criteria for model assessment. Section~\ref{sec:MCMCAlgo} describes the MCMC algorithm for our proposed model. A simulated example is presented in Section~\ref{sec:Sim}, illustrating the effectiveness of our modeling approach. Section~\ref{sec:App} demonstrates the application of our methodology through a stopover duration analysis for our motivating data collected by the Ottenby Bird Observatory in Sweden. Discussion is provided in Section~\ref{sec:Discu}. Further details surrounding the full conditional distributions and the MCMC sampling algorithm are provided in a Supplementary Appendix.

\section{The Mallard Data}\label{sec:data}
      The mallard (\textit{Anas platyrhynchos}), is the most common and widespread dabbling duck in the Northern hemisphere and an important model species for studies of ecological processes \citep{gunnarsson2012disease}, harvest management \citep{nichols2007adaptive}, and epidemiology of bird borne infections \citep{latorre2009effects}. It is a partial migrant, where southernly populations in the distribution range tend to be resident and the northernmost obligatory migrants, and in other populations a mix of resident and migrants \citep{cramphandbook1977}. The mallard is a medium-sized bird with heavy wing loading where migration is energetically costly. From ringing and telemetry studies it is clear that migratory mallards break up their journey into shorter flights and spend a large proportion of their migration time at stopover sites, replenishing resources and preparing for the next leg of migration \citep{gunnarsson2012disease}. Thus, stopover sites have a key role for successful migration and survival of mallards, and a priority for sustainable mallard management is to better characterize the ecology of birds at stopover. This includes assessing the timing of migration and densities of birds at specific stopover sites and to what extent intrinsic and extrinsic factors (e.g., body condition and weather) affect stopover behaviors.
      
      In birds, fat is the main fuel for migration and it remains to be known how mallards adjust their stopover behavior and departure according to their refueling rates at the stopover site and their current body condition in terms of fat loads \citep{berthold2001bird}. In addition, weather is known to be linked with bird migration during departure but also aloft. In general, birds prefer initiating a flight when winds provide flight assistance, i.e., tailwinds, and under other conditions favorable for flying, such as under low rainfalls \citep{berthold2001bird}. Furthermore, understanding how and when mallards use stopover sites is a key step in forecasting avian influenza dynamics at these sites \citep{gunnarsson2012disease}.

\begin{figure}[h]
   \centering
   \includegraphics[scale=0.25]{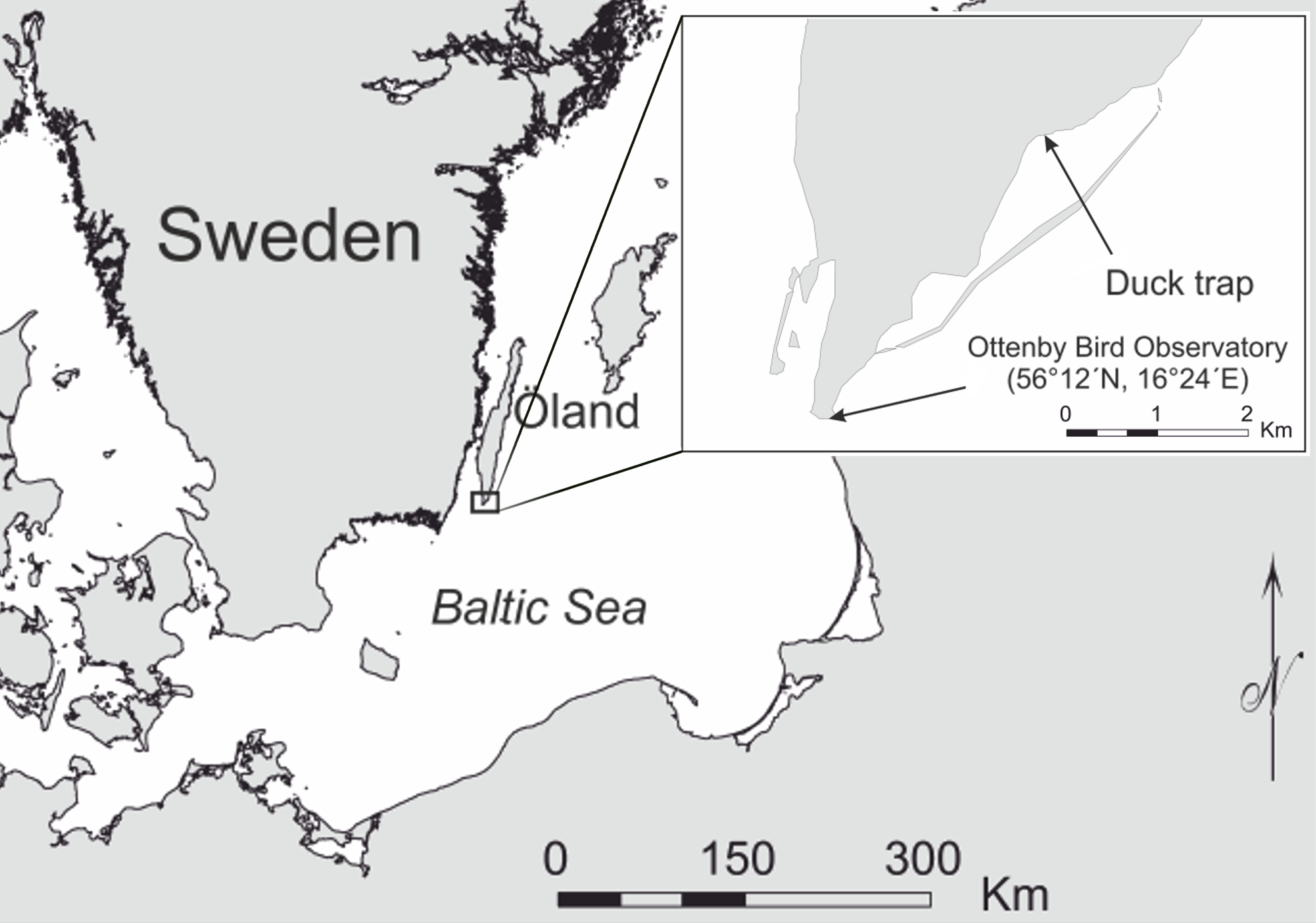}
   \caption{\baselineskip=10pt Plot of study site for monitoring mallards in the swedish island of \"{O}land (Section \ref{sec:data}).} 
   \label{fig:mallard_studysite}
\end{figure}

      Despite their importance in research, a lot remains to be determined in regards to mallard migration ecology, especially during the less well-studied stopover periods. Key objectives for monitoring studies of mallards---and indeed for other migratory birds more generally---are to understand stopover retention probabilities, stopover duration, total stopover population size (i.e., the total number of individuals present) at specific sites, and the effects of intrinsic and extrinsic factors on migratory decisions and stopover behaviors. Here we use long-term capture series of mallards carried out at Ottenby Bird Observatory on the Swedish island of \"{O}land in the Baltic Sea ($56^\circ 12'$N, $16^\circ 24'$E) (see Figure~\ref{fig:mallard_studysite}). This scheme started in 2002, and originally aimed for monitoring presence of influenza A virus in birds, but the data of banded individuals over time is also very suitable for addressing stopover ecology questions. The southernmost part of this island is an attractive stopover site for mallards within the Northwest European flyway, offering habitats for both roosting and foraging \citep{bengtsson2014movements}. Mallards that utilize our study site---Ottenby---mainly originate from mainland Sweden, Estonia, Finland, and Russia \citep{gunnarsson2012disease}. After leaving \"{O}land, these mallards migrate to wintering areas in Northwestern Europe, predominantly in southern Denmark, northern Germany, and the Netherlands \citep{gunnarsson2012disease}. 

      To collect data, Ottenby Bird Observatory used a stationary trap at the study site to catch mallards for ringing and epidemiological studies. In particular, mallards were attracted by bait grain and by the presence of a few (normally around 10) domestic ducks kept in a compartment of the trap. Traps were inspected daily during the field seasons and any wild duck captured was ringed and measured for structural size (i.e., the distance from the tip of the bill to the back of the head) and body mass, and subsequently released. This data collection process, over the course of a stopover season, results in the capture-recapture data. The data available for analysis was collected from 2004-2011, during the autumn migration season, which begins on August 1st and ends on December 16th of each year.

      Motivated by the mallard data at hand, our primary goal is to develop a model that accomplishes three important research objectives. The first objective is to determine whether there is individual heterogeneity in mallards' departure due to the intrinsic factor---body condition (i.e., body mass corrected by the structural size). The second objective is to estimate stopover duration, daily stopover population sizes, and total stopover population size, as well as to detect whether there is a temporal trend for daily stopover population sizes. The third objective is to understand how extrinsic factors such as wind and temperature relate to the timing of arrival and departure for mallards at our study site.

\section{Methodology}\label{sec:methods}
\subsection{Parameters and Notation}
     Consider a capture-recapture experiment with $T$ sampling occasions at distinct times $t_{1},t_{2},\ldots, t_{T}$ studying a population $\mcal{P}$ regarding a particular species of interest. Further, we assume the population size for population $\mcal{P}$ during the study is $N$, an unknown parameter that needs to be estimated. For $k=1,2,\ldots,T-1$, let $\delta_{k}=t_{k+1}-t_{k}$ denote the time interval between two consecutive sampling occasions $k$ and $k+1$. Without loss of generality, we assume $t_{1}<t_{2}<\cdots<t_{T}$; i.e., $\delta_{k}>0$ for $k=1,2,\ldots,T-1$. In addition, let $n$ be the total number of individuals that are caught during the study. For each individual being caught, denote $\mbv{y}_{i}=(y_{i,1},y_{i,2},\ldots,y_{i,T})$ as the corresponding capture history, where $y_{i,t}$ is a binary variable indicating if individual $i$ is caught at occasion $t$ for $i=1,2,\ldots,n$ and $t=1,2,\ldots,T$; that is, $y_{i,t}=1$ if individual $i$ is caught at occasion $t$ and 0 otherwise.  Upon the capture of each individual animal, measurements on a set of individual covariates are taken and recorded. 
     
     Motivated by the mallard data, we consider the single covariate case and allow the individual covariate $X_{i,t}$ to be continuous and time-varying. In the current context, we emphasize that the values of such a covariate for an individual are observable only when the individual is captured. As a result, we need to model the evolution of the time-varying continuous individual covariate. 
      
\subsection{Modeling Continuous Covariates}\label{sec:OU} 
       Let $X(t)$ be a continuous variable at time $t \in \mcal{T}=[0,T]$. We assume that $X(t)$ follows an Ornstein-Uhlenbeck (OU) process; i.e., $X(t)$ satisfies a stochastic differential equation of the form        
       \begin{equation}
             dX(t)=\tau(\alpha-X(t))dt+\sigma dW(t),
             \label{eq:growth_sde}
       \end{equation} 
       where $\sigma>0$ controls the noise variance, $\tau>0$ describes the rate of mean reversion, $\alpha$ is the long-term (or asymptotic) mean, and $W(t)$ is a standard Wiener process on $t\in \mcal{T}$. It is straightforward to see that by setting $\sigma=0$, (\ref{eq:growth_sde}) reduces to the von Bertalanffy growth equation \citep{von1938quantitative}. The use of the OU process in the current context is advantageous. The extra random noise term in the OU process provides increased flexibility, accounting for random noise resulting from several factors, e.g., measurement error and/or random variation due to changes in the environmental conditions \citep{filipe2010modelling}. For $t\in \mcal{T}$ and denote $X_{t}=X(t)$, the OU process is {\it stationary} (i.e., $(X_{t_{1}},X_{t_{2}},\ldots,X_{t_{s}})$ and $(X_{t_{1}+h},X_{t_{2}+h},\ldots,X_{t_{s}+h})$ are identically distributed),  {\it Markovian} (i.e., $P(X_{t_{s}}\leq x|X_{t_{1}},X_{t_{2}},\ldots,X_{t_{s-1}})=P(X_{t_{s}}\leq x|X_{t_{s-1}})$), and $(X_{t_{1}},X_{t_{2}},\ldots,X_{t_{s}})$ follows a multivariate {\it Gaussian} distribution \citep[see][and the references therein]{finch2004ornstein} for $t_{1}<t_{2}<\cdots <t_{s}$ and $h>0$. 
\noindent 

       The two moments of the OU process are: $E(X_{t})=\alpha$ and $\mbox{Cov}(X_{t},X_{s})=\sigma^{2}/(2\tau)\exp\{-\tau|s-t|\}$. For $t_{k-1}<t_{k}$, it follows that the transition distribution takes the following form
       \begin{equation*}
             X(t_{k})|\alpha,\tau,\sigma^{2},X(t_{k-1})=x_{k-1}\sim \mbox{N}\left(\mu(x_{k-1},\tau,\delta_{k-1},\alpha),V(\sigma^{2},\tau,\delta_{k-1})\right)
       \end{equation*}
       where
       \begin{align*}
             \mu(x_{k-1},\tau,\delta_{k-1},\alpha)&=\exp(-\tau \delta_{k-1})x_{k-1}+\left\{1-\exp(-\tau \delta_{k-1})\right\}\alpha \\
             V(\sigma^{2},\tau,\delta_{k-1})&=\frac{\sigma^{2}\left\{1-\exp(-2\tau \delta_{k-1})\right\}}{2\tau},
       \end{align*}
       \citep[see][and the references therein]{filipe2010modelling}. Compared with the diffusion process used by \citet{bonner2006extension,bonner2009JSIndCov} and \citet{schofield2011full}, the OU process we consider provides estimates for the rate parameter $\tau$ and long-term mean $\alpha$.
             
       For $i\in \mcal{P}$ and $t=1,2,\ldots,T$, the time-varying continuous individual covariate $X_{i,t}$ is assumed to satisfy the OU process defined by (\ref{eq:growth_sde}). Hence, at discrete sampling times $t=2,\ldots,T$, the conditional distribution for $X_{i,t}$ takes the following form
       \begin{equation*}
             X_{i,t}|\alpha,\tau,\sigma^{2},X_{i,t-1}=x_{i,t-1}\sim \mbox{N}\left(\mu(x_{i,t-1},\tau,\delta_{t-1},\alpha),V(\sigma^{2},\tau,\delta_{t-1})\right),
       \end{equation*}
       where $x_{i,t}$ is the realization of $X_{i,t}$ and $X_{i,1}\stackrel{iid}\sim \mbox{N}(x_{0},\sigma^{2}_{0})$.

\subsection{Semiparametric Jolly-Seber Model with Individual Heterogeneity}  

    The JS model we propose is formulated under the state-space framework. In particular, our proposed model is characterized by a state model, observation model, and parameter model. The state model describes the states of an individual over time, whereas the observation model describes the capture outcome of an individual over time. Throughout this article, we use the term ``state" to describe two statuses of an individual in the population, which are either alive and present in the study area (denoted by 1) or not having entered the population or death (denoted by 0). The parameter model describes how certain model parameters are linked to the intrinsic and extrinsic factors. 
    
    Let $\mbv{z}_{i}=(z_{i,1},z_{i,2},\ldots,z_{iT})$ where $z_{i,t}$ is a binary latent variable to indicate the state of individual $i$ at time $t$ for $1 \leq i \leq N$ and $1 \leq t \leq T$. Note that the dimension of $\mbf{z}=\{\mbv{z}_{1},\mbv{z}_{2},\ldots,\mbv{z}_{N}\}$ varies with $N$, a parameter that is unknown. Consequently, the number of parameters is not fixed in each iteration of MCMC, which will cause some computational disadvantages. To maintain a constant number of parameters, a \textit{data augmentation} technique is often utilized \citep[e.g., see][]{royle2008hierarchical}. For our model, the data augmentation technique involves two steps. The first step is to introduce a parameter $M>N$, and augment the observed data configuration $\mbf{y}_{obs}=\{\mbv{y}_{1},\mbv{y}_{2},\ldots,\mbv{y}_{n}\}$ by $\mbf{y}_{\text{aug}}=\{\mbv{y}_{n+1},\mbv{y}_{n+2},\ldots,\mbv{y}_{N},\ldots,\mbv{y}_{M}\}$, where $\mbv{y}_{i}=\mbv{0}$ for $i=n+1,n+2,\ldots,M$. Second, for $i=1,2,\ldots,M$, we associate a binary membership indicator $w_{i}$ with each of $M$ individuals; i.e., $w_{i}\stackrel{iid}\sim \text{Bernoulli}(\Psi)$. In other words, $w_{i}=1$ if individual $i$ is a member of $\mcal{P}$ and 0 otherwise. 
  
\subsubsection{State Model}  
    Following \citet{royle2008hierarchical}, the \textit{state model} can be defined by
    \begin{eqnarray}
          z_{i,1}|w_{i},\pi_{1}^{c}&\sim& \text{Bernoulli}(w_{i}\pi_{1}^{c}),  \label{eq:state_t1} \\
          z_{i,t+1}|z_{i,t},\phi_{i,t},w_{i},\pi_{t+1}^{c}&\sim& \text{Bernoulli}(\phi_{i,t}z_{i,t}+w_{i}\pi_{t+1}^{c}R_{i,t}), \label{eq:state_t2}
    \end{eqnarray}
    where $R_{i,t}=\prod_{s=1}^{t}1(z_{i,s}=0)$ indicates whether an individual $i$ can enter the population right after time $t$ for $i=1,2,\ldots,M$ and $t=1,2,\ldots,T-1$. In addition, $1(z=a)$ is the indicator function that takes value 1 if $z=a$ and 0 otherwise. For $t=1,2,\ldots,T-1$, $\phi_{i,t}$ refers to survival probability (or stopover retention probability in a stopover model), i.e., the probability that an individual $i$ of $\mcal{P}$ will remain in the study area at time $t+1$ given its presence in the study area at time $t$. Moreover, $\pi_{t}^{c}$ denotes the conditional entrance probability at time $t$ given that an individual has not entered the study area, that is,
    \begin{equation*}
         \pi_{t+1}^{c}=\frac{\beta_{t}}{\sum_{j=t}^{T-1}{\beta_{j}}},
   \end{equation*}
for $t=0,1,\ldots,T-1$ and $\beta_{t}$ denotes the proportion of $\mcal{P}$ that enters the study area between time $t$ and $t+1$. By definition, it follows that $\sum_{t=1}^{T}{\beta_{t-1}}=1$.
    
    The interpretation of the state model described in (\ref{eq:state_t1}) and (\ref{eq:state_t2}) is straightforward. First, (\ref{eq:state_t1}) indicates that individual $i$ is subject to entrance with probability $\pi_{1}^{c}$ at time $t=1$ only if it is a member of $\mcal{P}$ (i.e., $w_{i}=1$). In (\ref{eq:state_t2}), we see that if individual $i$ has not entered the study area right before time $t+1$ (i.e., $R_{i,t}=1$), it is subject to entrance with probability $\pi_{t+1}^{c}$ given it is a member of $\mcal{P}$. Second, if individual $i$ has already entered and is present in the study area at time $t$, it will remain in the study area at time $t+1$ with probability $\phi_{i,t}$. 
  
\subsubsection{Observation Model}  
    We proceed with the \textit{observation model}. For $1 \leq t \leq T$, denote $p_{t}$ as the capture probability at time $t$. The observation model is given by
      \begin{equation}
            y_{i,t}|w_{i},z_{i,t},p_{t}\sim \text{Bernoulli} (w_{i}z_{i,t}p_{t}),
            \label{eq:observation}
      \end{equation}
for $i=1,2,\ldots,M$. According to (\ref{eq:observation}), we are solely interested in the capture outcome for individuals that are members of $\mcal{P}$ (i.e., for any $i=1,2,\ldots,M$ such that $w_{i}=1$). Moreover, for individual $i$ that is captured at least once during the study (i.e., $\mbv{y}_{i}\neq \mbv{0}$), it is clear that $w_{i}=1$ is implied. In addition, individual $i$ is subject to capture at time $t$ only if it has entered and still remains in the study area (i.e., $z_{i,t}=1$).

    An important feature of building the JS model from the ``individual" up is that it enables us to estimate certain quantities that are important in stopover duration analysis fairly easily. For example, the total stopover population size, $N$, can be estimated as $N=\sum_{i=1}^{M}w_{i}$. The stopover population size at time $t$, $N_{t}$,  can be estimated as $N_{t}=\sum_{i=1}^{M}w_{i}z_{i,t}$. Moreover, we can estimate the mean stopover duration averaged over all captured individuals as \citep{lyons2015population}
    \begin{equation*}
          S=\frac{\sum_{i=1}^{n}\sum_{t=1}^{T}z_{i,t}}{n}.
    \end{equation*}
The number of individuals alive at both times $t_{1}$ and $t_{2}$, say $N_{t_{1},t_{2}}$, can be calculated as $N_{t_{1},t_{2}}=\sum_{i=1}^{M}z_{i,t_{1}}z_{i,t_{2}}w_{i}$.

\subsubsection{Parameter Model}
      The \textit{parameter model} links capture, departure, and entrance parameters with various types of covariates. We consider a semiparametric model for the capture probabilities. The departure probabilities are linked to a time-varying continuous individual covariate to account for individual heterogeneity. Additionally, we consider a model that links the entrance probabilities to time dependent covariates to infer the impacts of these covariates on the timing of arrival.   
      
      Starting with capture probabilities $p_{t}$, we consider a semiparametric model of the form
      \begin{equation}
         \mbox{logit}(p_{t})=\mbv{g}_{t}^{\prime}\mbv{\zeta}+\sum_{k=1}^{K}u_{k}|o_{t}-\kappa_{k}|^{3},
         \label{eq:pcap_semipar_orig}
      \end{equation}
where $\mbox{logit}(r)=\log\{r/(1-r)\}$ and $K$ is the number of knot points. Here $\mbv{g}_{t}=(g_{1t},g_{2t},\ldots,g_{Qt})^{\prime}$ is a $Q\times 1$ vector consists of values for covariates $g_{1}, g_{2}, \ldots, g_{Q}$ at time $t$; and $\mbv{\zeta}=(\zeta_{1},\zeta_{2},\ldots,\zeta_{Q})^{\prime}$ denotes a $Q\times 1$ vector of regression coefficients. Moreover, it is assumed that $\mbv{u}=(u_{1},u_{2},\ldots,u_{K})^{\prime}\sim N(0,\sigma_{u}^{2}\mbf{\Omega}^{-1})$ where $\mbf{\Omega}$ is a matrix whose $(k,l)$th entry is $|\kappa_{k}-\kappa_{l}|^{3}$ for $1\leq k,l \leq K$. Following \citet{ruppert2003semiparametric}, the fixed knot $\kappa_{k}$ is chosen to be sample quantile of the $o_{t}$'s corresponding to probability $k/(K + 1)$ for $k=1,2,\ldots,K$ where $K=\max\left\{20,\min\left(150,\frac{T}{4}\right)\right\}$. Let $\mbf{Z}_{K}$ be the matrix with $t$th row $\mbv{Z}_{Kt}=(|o_{t}-\kappa_{1}|^{3},|o_{t}-\kappa_{2}|^{3},\ldots,|o_{t}-\kappa_{K}|^{3})$, (\ref{eq:pcap_semipar_orig}) can be reparameterized as
      \begin{equation}
         \mbox{logit}(p_{t})=\mbv{g}_{t}^{\prime}\mbv{\zeta}+\mbv{Z}_{t}\mbv{b},
         \label{eq:pcap_semipar}
      \end{equation}
      where $\mbv{b}=\mbf{\Omega}^{\frac{1}{2}}\mbv{u}$ and $\mbv{Z}_{t}$ is the $t$th row of the matrix $\mbf{Z}=\mbf{Z}_{K}\mbf{\Omega}^{\frac{1}{2}}$. Due to this reparameterization, it holds that 
$\mbv{b}\sim N(\mbv{0},\sigma_{u}^{2}\mbf{I}_{K})$ where $\mbf{I}_{K}$ is a $K\times K$ identity matrix.
      
      From a modeling perspective, the parametric part of (\ref{eq:pcap_semipar}) posits a linear relationship between covariates $g_{1},g_{2},\ldots, g_{Q}$ and the logit of $p_{t}$. In comparison, the nonparametric part of (\ref{eq:pcap_semipar}) allows for a greater flexibility in the sense that the shape of the functional relationship between the covariate $o$ and the logit of $p_{t}$ is determined by the data instead of assuming a particular parametric form a priori. For the nonparametric part of the model in (\ref{eq:pcap_semipar}), we consider low-rank thin-plate splines due to their appealing numerical properties in Bayesian computation. That is, the parameters associated with low-rank thin-plate splines tend to be less correlated than parameters associated with other basis functions, which leads to better mixing of the MCMC chains in Bayesian analysis \citep{crainiceanu2007bayesian}. 
      
      Define $d_{i,t}=1-\phi_{i,t}$ as the departure probability of individual $i\in \mcal{P}$ at time $t$ for $t=1,2,\ldots,T-1$. Strictly speaking, departures can arise from three outcomes---start of a migratory flight, relocation to another habitat that is not covered by traps, and death. When the sampling period is relatively short, as it is the case in our motivating mallard example, death between two consecutive sampling periods is almost negligible. As a result, the term departure primarily refers to start of another migratory flight or relocation to another habitat. We link $d_{i,t}$ to an intrinsic factor $X$ as follows
      \begin{equation}
         \mbox{logit}(d_{i,t})=\mbv{x}_{i,t}'\mbv{\theta}=\theta_{0}+\theta_{1}x_{i,t}.
         \label{eq:surv_param} 
      \end{equation}
Here the realization of a time-varying continuous individual covariate (i.e., $\{x_{i,t}\}$) accounts for individual heterogeneity in departure. As previously mentioned, the inclusion of a time-varying continuous individual covariate raises some computational concerns. First, for an individual $i$ that is not captured at time $t$, the value of $x_{i,t}$ is not observable. Further, for individuals that are never captured during the study, we do not observe any values for $x_{i,t}$. Accordingly, the implementation of the JS model we propose requires us to establish a model for the covariate such that missing values can be ``imputed" by conditioning on the observed data. To achieve this goal, we assume $X_{i,t}$ follows the OU process discussed in Section~\ref{sec:OU}.

      For entrance probabilities, we consider the following model
      \begin{equation}
         \log \left(\frac{\beta_{t}}{\beta_{T-1}}\right)=\mbv{\Lambda}_{t}^{\prime}\mbv{\gamma},
         \label{eq:bent_param}
      \end{equation}
where $\mbv{\Lambda}_{t}=(\Lambda_{1t},\Lambda_{2t},\ldots,\Lambda_{Pt})^{\prime}$ denotes a $P\times 1$ vector consists of the values of covariates $\Lambda_{1},\Lambda_{2},\ldots,\Lambda_{P}$ at time $t+1$ for $t=0,1,\ldots,T-2$. Furthermore, $\mbv{\gamma}=(\gamma_{1},\gamma_{2},\ldots,\gamma_{P})^{\prime}$ is a $P\times 1$ vector of regression coefficients. Due to the implied restriction $\sum_{t=1}^{T}\beta_{t-1}=1$, (\ref{eq:bent_param}) is equivalent to the following
      \begin{equation*}
         \beta_{t}=
         \begin{cases}
           \frac{\exp\left(\mbv{\Lambda}_{t}^{\prime}\mbv{\gamma}\right)}{1+\sum_{j=0}^{T-2}\exp\left(\mbv{\Lambda}_{j}^{\prime}\mbv{\gamma}\right)} & \mbox{if} \,\, t=0,1,\ldots,T-2 \\
           \frac{1}{1+\sum_{j=0}^{T-2}\exp\left(\mbv{\Lambda}_{j}^{\prime}\mbv{\gamma}\right)} & \mbox{if} \,\, t=T-1.
         \end{cases}
      \end{equation*}
          
 \subsection{Priors and Posteriors}
 
         To complete the specification of our model, we need to assign prior distributions for the model parameters and derive the full conditional distributions. Denote $\mbv{w}=\{w_{i}: i=1,2,\ldots,M\}$, the set of parameters in the model we propose is $\mbv{\Theta}=\{\mbv{\zeta},\mbv{b},\mbv{\theta},\mbv{\gamma},\Psi,\mbv{w},\mbf{z},x_{0},\sigma_{0}^{2},\alpha,\tau,\sigma^{2},\sigma_{u}^{2}\}$.  Denote $\mbox{IG}(A,B)$ as the inverse gamma distribution with shape parameter $A$ and scale parameter $B$, we assign prior distributions as follows: $\mbv{\zeta} \sim \mbox{N}(\mbv{\mu}_{\zeta},\mbf{\Sigma}_{\zeta})$; $\mbv{b}\sim \mbox{N}(\mbv{0},\sigma_{b}^{2}\mbf{I}_{K})$; $\mbv{\theta} \sim \mbox{N}(\mbv{\mu}_{\theta},\mbf{\Sigma}_{\theta})$; $\mbv{\gamma} \sim \mbox{N}(\mbv{\mu}_{\gamma},\mbf{\Sigma}_{\gamma})$; $w_{i}\stackrel{iid}\sim \mbox{Bernoulli}(\Psi)$ for $i=1,2,\ldots,M$; $\Psi \sim \mbox{Beta}(a_{\Psi},b_{\Psi})$; $x_{0}\sim \mbox{N}(\mu_{0},\sigma_{x_{0}}^{2})$; $\sigma_{0}^{2}\sim \mbox{IG}(q_{\sigma_{0}},r_{\sigma_{0}})$; $\alpha \sim \mbox{N}(\mu_{\alpha},\sigma_{\alpha}^{2})$; $\tau \sim \mbox{Unif}(q_{\tau},r_{\tau})$; $\sigma^{2} \sim \mbox{IG}(q_{\sigma},r_{\sigma})$; and $\sigma_{u}^{2} \sim \mbox{IG}(q_{u},r_{u})$. In our implementation, we choose vague priors that are noninformative relative to the scale of data. 
          
         Let $\mbf{Y}=\mbf{y}_{\text{obs}}\cup \mbf{y}_{\text{aug}}$ denote the observed capture history. Assuming conditional independence, the joint posterior distributions of the model parameters $[\mbv{\Theta}|\mbf{Y}]$ can be derived as
         \begin{align*}
            [\mbv{\Theta}|\mbf{Y}]&\propto\left\{\prod_{i=1}^{M}\left(\prod_{t=1}^{T-1}[z_{i,t+1}|z_{i,t},w_{i},\mbv{\gamma},x_{i,t},\mbv{\theta}][x_{i,t+1}|x_{i,t},\alpha,\tau,\sigma^{2}]\right)[z_{i,1}|w_{i},\mbv{\gamma}][x_{i,1}|x_{0},\sigma_{0}^{2}] \right.\notag \\
             &\times \left. \left(\prod_{t=1}^{T}[y_{i,t}|z_{i,t},w_{i},\mbv{\zeta},\mbv{b}]\right)[w_{i}|\Psi]\right\}[\mbv{\theta}][\tau][\alpha][\sigma^{2}][x_{0}][\sigma_{0}^{2}][\mbv{\gamma}][\mbv{b}|\sigma_{u}^{2}][\Psi][\sigma_{u}^{2}].
            %\label{eq:joint_post}
         \end{align*}
    
\subsection{Model Assessment}
       An extremely important aspect of Bayesian modeling is to evaluate goodness-of-fit for the model being considered. In the context of capture-recapture models, the Bayesian p-value is often considered \citep[e.g., see][and the references therein]{king2010bayesian}.  Roughly speaking, the Bayesian p-value is a posterior probability that measures the similarity between the data generated from the posterior predictive distribution under a specified model and the observed data. To calculate the Bayesian p-value, we first define a discrepancy function $h(\mbv{D},\mbv{\Theta})$, where $\mbv{D}$ and $\mbv{\Theta}$ denote the data and the parameters for the model being considered, respectively. Then, we calculate the value of the discrepancy function for both the observed data $\mbv{D}^{\star}$ and the simulated data $\mbv{D}'$, which is generated conditioning on the posterior distribution of model parameters. Finally, the Bayesian p-value is defined as the percentage of times that values of the discrepancy function for $\mbv{D}^{\star}$ exceeds those of the discrepancy function for $\mbv{D}'$. Mathematically, the definition of the Bayesian p-value, $P_{b}$, can be formulated as $P_{b}=p(h(\mbv{D}^{\star},\mbv{\Theta})>h(\mbv{D}',\mbv{\Theta})|\mbv{D}^{\star})$. As a rule of thumb, a Bayesian p-value close to 0 or 1 indicates that the model being considered does not provide a good fit to the data and that there is inconsistence between the model and data \citep[see Chapter 6 in][]{gelman2003bayesian}.

       For the model we propose, goodness-of-fit requires the assessment of two components. On the oned hand, we need to assess the goodness-of-fit for the overall JS model to the data. On the other hand, we need to evaluate the use of the OU process regarding modeling the time-varying continuous individual covariate. Consequently, it suffices to calculate the Bayesian p-values $P_{b}^{JS}$ for the JS model and $P_{b}^{OU}$ for modeling the individual covariate using the OU process. Among many choices of the discrepancy function \citep[e.g., see][]{brooks2000bayesian}, we used the complete log-likelihood function for $P_{b}^{JS}$; i.e., $h^{JS}(\mbv{D},\mbv{\Theta})=\ell(\mbf{Y},\mbv{z}|\mbv{\Theta}_{-\mbv{z}},\mbv{D})$, where $\ell(\mbf{Y},\mbv{z}|\mbv{\Theta},\mbv{D})$ is the complete log-likelihood function of $\mbf{Y},\mbv{z}$ given all model parameters excluding $\mbv{z}$ (i.e., $\mbv{\Theta}_{-\mbv{z}}$) and the data $\mbv{D}$. Different from \cite{bonner2009JSIndCov}, for $P_{b}^{OU}$, we compare the observed and expected value of the individual covariate for each capture rather than recapture and consider the discrepancy function to be
        \begin{equation*}
            h^{OU}(\mbv{D},\mbv{\Theta})=\frac{1}{n_{c}}\sum_{i\in \mcal{P}}\sum_{t: y_{i,t}=1} \left\{\frac{x_{i,t}-E(x_{i,t}|x_{i,t-1})}{\sigma(x_{i,t}|x_{i,t-1})}\right\}^{2},
       \end{equation*} 
 where $n_{c}=\sum_{i\in \mcal{P}}\sum_{t=1}^{T}y_{i,t}$ is the total number of captures over $T$ sampling occasions and $\sigma(x_{i,t}|x_{i,t-1})$ denotes the standard deviation for the distribution of $x_{i,t}|x_{i,t-1}$.

\section{MCMC Algorithm}\label{sec:MCMCAlgo}
   We describe our customized MCMC sampling algorithms for $\mbv{z}$, $\mbv{\zeta}$, $\mbv{b}$, and $\mbv{w}$. For the rest of model parameters, the details are provided in the Supplementary Appendix.
  
\subsection{Sampling $\mbv{z}$} 
      We now discuss how to update the latent variables $\mbv{z}$. For $i=1,2,\ldots,M$, we first define three sets as follows:
      \begin{align*}
            S_{1}&=\{i: w_{i}=0\} \\
            S_{2}&=\{i: \mbv{y}_{i}\ne \mbv{0}\} \\
            S_{3}&=\{i: \mbv{y}_{i}=\mbv{0}, w_{i}=1\}.
      \end{align*}
      The update of $\mbv{z}_{i}$ will depend on which category an individual $i$ falls into. For example, if an individual $i$ is not a member of $\mcal{P}$, i.e., $i \in S_{1}$, we always fix $\mbv{z}_{i}=\mbv{0}$. Second, for an individual  $i \in S_{2}$, it is captured at least once during the $T$ sampling occasions. As a consequence, $y_{i,t}=1$ would necessarily imply $z_{i,t}=1$ for $i \in S_{2}$, since an individual needs to be alive and present in the study area in order to be available for capture. In this case, the simulation of $\mbv{z}_{i}$ depends on the structure of $\mbv{y}_{i}$. Consider a capture history of the form
       \begin{equation}
             \mbv{y}_{i}=00010100
             \label{eq:caphist_example}
       \end{equation} 
       with $T=8$. It is clear that the corresponding latent states $\mbv{z}_{i}$ takes the form of $\mbv{z}_{i}=\cdot\cdot\cdot111\cdot\cdot$, where $\cdot$ denotes missing states to be simulated. 

      We start with the updating scheme of $\mbv{z}_{i}$ for $i \in S_{2}$. To simplify notation, we denote $f_{i}$ and $l_{i}$ as the first and last times that an individual $i$ is captured. We adopt a block updating scheme similar to \cite{dupuis2007bayesian}. Specifically, let $B_{1}(i)$ be the Type I block that consists of state variables corresponding to sample times up to $f_{i}$.  Further, denote $B_{2}(i)$ as the Type II block that consists of state variables corresponding to sampling occasions after $l_{i}$. For example, for the capture history in (\ref{eq:caphist_example}), we have $B_{1}(i)=\{z_{i,1},z_{i,2},z_{i,3},z_{i,4}\}$ and $B_{2}(i)=\{z_{i,7},z_{i,8}\}$. Before we proceed with the simulation for Type I and Type II blocks, we need to introduce some further notation. Let $\lambda_{i,t}$ denote the probability that individual $i$ enters the population, is still alive, and is not seen before time $t$, the following recursive relationship holds 
     \begin{equation*}
            \lambda_{i,t+1}=\beta_{t}+\lambda_{i,t}(1-p_{t})\phi_{i,t}
     \end{equation*}
for $t=1,2,\ldots,T-1$ and $\lambda_{i,1}=\beta_{0}$. Consequently, for Type I block $B_{1}(i)$, we can update $B_{1}(i)=(z_{i,1},\ldots,z_{i,f_{i}})$ according to $B_{1}(i) \sim \mbox{Multinomial}(1,\mbv{\xi}_{i})$ where $\mbv{\xi}_{i}=(\xi_{i,1},\xi_{i,2},\ldots,\xi_{i,f_{i}})$ and
    \begin{equation*}
          \xi_{i,t}=\frac{\beta_{t-1}\prod_{s=t}^{f_{i}-1}(1-p_{s})\phi_{i,s}}{\lambda_{i,f_{i}}}
    \end{equation*}
    for $t=1,2,\ldots,f_{i}$.
    
     Next, we discuss the simulation for latent state variables in the Type II block. Let $v_{i,t}$ denote the probability that an individual $i$ of $\mcal{P}$ leaves the study area after time $t$, we can then obtain $v_{i,t}$ using the recursion
     \begin{equation*}
           v_{i,t}=1-\phi_{i,t}+\phi_{i,t}(1-p_{t+1})v_{i,t+1}
    \end{equation*}
    for $t=T-1,T-2,\ldots,1$ and $v_{i,T}=1$. Accordingly, for $t=l_{i}+1,\ldots,T$, we can update $z_{i,t}\in B_{2}(i)$ by first simulating $\eta_{i,t}$ from
    \begin{equation*}
       \eta_{i,t}\sim\text{Bernoulli}\left(\frac{1-\phi_{i,t-1}}{v_{i,t-1}}\right)
    \end{equation*} 
    and then update $z_{i,t}$ according to
    \begin{equation*}
          z_{i,t}=
          \begin{cases}
              1             & \mbox{if $z_{i,t-1}=1$ and $\eta_{i,t}=0$} \\
              0             & \mbox{otherwise}.
          \end{cases}
    \end{equation*}
 
   Lastly, we address the simulation of latent state variables $\mbv{z}_{i}$ for an individual $i$ of $\mcal{P}$ that is never captured during the entire study (i.e., $i \in S_{3}$). To achieve this goal, let $\varrho_{i}$ denote the probability that individual $i$ of $\mcal{P}$ is never captured. We can derive the following
      \begin{equation}
         \varrho_{i}=1-\sum_{t=1}^{T}\lambda_{i,t}p_{t}.
         \label{eq:rhoeq}
      \end{equation}
 To perform Type I block simulation, we first determine the time that individual $i$ of $\mcal{P}$ first enters the population according to $\mbv{z}_{i} \sim \mbox{Multinomial}(1,\mbv{\iota}_{i})$ with $\mbv{\iota}_{i}=(\iota_{i,1},\iota_{i,2},\ldots,\iota_{i,T})$ and
       \begin{equation*}
             \iota_{i,t}=\frac{\beta_{t-1}(1-p_{t})v_{i,t}}{\varrho_{i}}
       \end{equation*}
       for $t=1,2,\ldots,T$. After determining the time of entrance into the population, we need to perform Type II block simulation to ascertain the status of individual $i$ after its entrance.  For the sake of brevity, the details are omitted here due to its similarity with the Type II block simulation for $i \in S_{2}$ in the previous discussion.

\subsection{Sampling $\mbv{\zeta}$ and $\mbv{b}$}
    Denote $U_{t}=\sum_{i=1}^{n}y_{i,t}$, the joint conditional distribution of $(\mbv{\zeta},\mbv{b})$ takes the form of
            \begin{equation}
               [\mbv{\zeta},\mbv{b}|\cdot]\propto \left\{\prod_{t=1}^{T}\frac{\exp\left(U_{t}(\mbv{g}_{t}^{\prime}\mbv{\zeta}+\mbv{Z}_{t}\mbv{b})\right)}{\left(1+\exp(\mbv{g}_{t}^{\prime}\mbv{\zeta}+\mbv{Z}_{t}\mbv{b})\right)^{N_{t}}}\right\}\exp\left(-\frac{(\mbv{\zeta}-\mbv{\mu}_{\zeta})^{\prime}\mbf{\Sigma}_{\zeta}^{-1}(\mbv{\zeta}-\mbv{\mu}_{\zeta})}{2}\right)\exp\left(-\frac{\mbv{b}^{\prime}\mbv{b}}{2\sigma_{u}^{2}}\right),
               \label{eq:zetab_cond}
            \end{equation}
which is not of standard form. To avoid tuning, we take advantage of the following results \citep{polson2013bayesian}
   \begin{equation}
      \frac{(e^{\psi})^{A}}{(1+e^{\psi})^{B}}=2^{-B}e^{(A-\frac{B}{2}) \psi}\int_{0}^{\infty} e^{-\frac{\omega \psi^{2}}{2}}\mbox{PG}(\omega|B,0) d\omega.
      \label{eq:PolyaGamma_binprob}
   \end{equation}
   Here $\mbox{PG}(\omega|C,D)$ denotes a P{\'o}lya--Gamma distribution with parameters $C>0$ and $D\in \mcal{R}$ and the corresponding probability density function being \citep{polson2013bayesian}:
   \begin{equation*}
      \mbox{PG}(\omega|C,D)=\frac{\exp\left(-\frac{D^{2}\omega}{2}\right)\mbox{PG}(\omega|C,0)}{E_{\omega}\left\{\exp\left(-\frac{D^{2}\omega}{2}\right)\right\}}.
   \end{equation*}  
  
  Combining (\ref{eq:zetab_cond}) and (\ref{eq:PolyaGamma_binprob}) yields $\mbv{\zeta}|\cdot \sim \mbox{N}(\widetilde{\mbv{\mu}}_{\zeta},\widetilde{\mbf{\Sigma}}_{\zeta})$ with
            \begin{align*}
               \widetilde{\mbf{\Sigma}}_{\zeta}&=\left(\mbf{G}^{\prime}\mbf{D}_{q}\mbf{G}+\mbf{\Sigma}_{\zeta}^{-1}\right)^{-1} \\
               \widetilde{\mbv{\mu}}_{\zeta}&=\widetilde{\mbf{\Sigma}}_{\zeta}\left(\mbf{G}^{\prime}(\mbv{\kappa}_{u,N}-\mbf{D}_{q}\mbf{Z}\mbv{b})+\mbf{\Sigma}_{\zeta}^{-1}\mbv{\mu}_{\zeta}\right),
            \end{align*}
            where $\mbf{D}_{q}=\mbox{diag}(q_{1},q_{2},\ldots,q_{T})$ and $\mbv{\kappa}_{u,N}=\left(U_{1}-\frac{1}{2}N_{1}, U_{2}-\frac{1}{2}N_{2}, \ldots, U_{T}-\frac{1}{2}N_{T}\right)^{\prime}$. Moreover, $q_{t}|\cdot \sim \mbox{PG}(N_{t},\mbv{g}_{t}^{\prime}\mbv{\zeta}+\mbv{Z}_{t}\mbv{b})$ for $t=1,2,\ldots,T$. Regarding the conditional distribution of $\mbv{b}$, we have $\mbv{b}|\cdot \sim \mbox{N}(\widetilde{\mbv{\mu}}_{b},\widetilde{\mbf{\Sigma}}_{b})$ with
            \begin{align*}
               \widetilde{\mbf{\Sigma}}_{b}&=\left(\mbf{Z}^{\prime}\mbf{D}_{q}\mbf{Z}+\frac{1}{\sigma_{u}^{2}}\mbf{I}_{K}\right)^{-1} \\
               \widetilde{\mbv{\mu}}_{b}&=\widetilde{\mbf{\Sigma}}_{b}\left\{\mbf{Z}^{\prime}\left(\mbv{\kappa}_{u,N}-\mbf{D}_{q}\mbf{G}\mbv{\zeta}\right)\right\},
            \end{align*}
where $\mbf{G}$ is a $T\times Q$ matrix whose $t$th row consists of $\mbv{g}_{t}^{\prime}$. Advantageously, by introducing another layer of data augmentation using P{\'o}lya--Gamma distribution random variates, the full conditional distributions for $\mbv{\zeta}$ and $\mbv{b}$ have a standard form.

\subsection{Sampling $\mbv{w}$}
      We describe the sampling algorithm for membership indicator $w_{i}$, $i=1,2,\ldots,M$. For individuals $i\in S_{2}$, it is straightforward to see that $w_{i}=1$, i.e., $\text{P}(w_{i}=1|\mbv{y}_{i}\neq \mbv{0})=1$. In other words, for individuals that are captured at least once during the study, they are members of $\mcal{P}$. For an individual $i$ that is never captured, i.e., $i\in S_{3}$, we can apply Bayes rule to arrive at:
      \begin{equation*}
         \epsilon_{i}=\mbox{P}(w_{i}=1|\mbv{y}_{i}=\mbv{0})=\frac{\Psi \rho_{i}}{\Psi \rho_{i}+(1-\Psi)}
      \end{equation*}
and hence, we can sample $w_{i}$ according to $w_{i}|\cdot \sim \mbox{Bernoulli}(\epsilon_{i})$.

\section{Simulated Example}\label{sec:Sim}      
      To evaluate the performance of our proposed model, we consider a simulated example where the exact model specification is chosen for illustration. For this simulation, we set $N=8,000$ and $T=77$.  For the parameters specific to the OU process, we set $x_{0}=-0.64$, $\sigma_{0}^{2}=1.37$, $\alpha=0.20$, $\tau=0.19$, $\sigma^{2}=0.36$, and $\delta_{k}\equiv 1.0$ (for $k=1,2,\ldots,T-1$). In terms of the model for departure probability $d_{i,t}$, we consider 
      \begin{equation*}
        \mbox{logit}(d_{i,t})=\theta_{0}+\theta_{1}x_{i,t},
      \end{equation*}
where $i=1,2,\ldots,N$, $t=1,2,\ldots,T$, and $\mbv{\theta}=(\theta_{0},\theta_{1})^{\prime}=(-1.8,0.3)^{\prime}$. In addition, $x_{i,t}$ is the realization of a time-varying continuous individual covariate satisfying the OU process with the aforementioned parameter specification. 

       For the model associated with the capture probabilities $p_{t}$, we consider
       \begin{equation*}
         \mbox{logit}(p_{t})=\mbv{g}_{t}^{\prime}\mbv{\zeta}+\mbv{Z}_{t}\mbv{b},
       \end{equation*}
where $\mbv{g}_{t}=(g_{1t},g_{2t},g_{3t})^{\prime}$ for $t=1,2,\ldots,T$; and $g_{1}$, $g_{2}$, and $g_{3}$ are three time dependent covariates. These three covariates are simulated according to $g_{1t}, g_{2t}, g_{3t}\stackrel{iid}\sim \mbox{N}(0,1)$ for $t=1,2,\ldots,T$. For the regression coefficients $\mbv{\zeta}=(\zeta_{1},\zeta_{2},\zeta_{3})^{\prime}$, we consider $\mbv{\zeta}=(1.0,-0.9,0.6)^{\prime}$. In addition, $\mbv{Z}_{t}$ is the $t$th row of matrix $\mbf{Z}=\mbf{Z}_{K}\mbf{\Omega}_{K}^{-\frac{1}{2}}$. Here $\mbv{Z}_{K}$ is the matrix with $t$th row $\mbv{Z}_{Kt}=(|o_{t}-\kappa_{1}|^{3}, |o_{t}-\kappa_{2}|^{3}, \ldots, |o_{t}-\kappa_{K}|^{3})^{\prime}$ for $o_{t}=\frac{t}{T}$ and $t=1,2,\ldots,T$; and $\mbf{\Omega}_{K}$ is a $K\times K$ matrix whose $(k,l)$th entry is $|\kappa_{k}-\kappa_{l}|^{3}$ for $1\leq k,l\leq K$. Moreover, the $k$th fixed knot $\kappa_{k}$ is chosen as the sample quantile of $\{o_{1},o_{2},\ldots,o_{T}\}$ corresponding to probability $\frac{k}{K+1}$ for $k=1,2,\ldots,K$. This particular simulation setup for capture probabilities ensures that the resulting encounter history is neither too dense or too sparse. We chose the number of knots according to $K=\max\left\{20,\min\left(150,\frac{T}{4}\right)\right\}$, which yields $K=20$. For $\mbv{b}=(b_{1},b_{2},\ldots,b_{K})^{\prime}$, we choose
$b_{k}\stackrel{iid}\sim \mbox{N}(0,\sigma_{u}^{2})$ with $\sigma_{u}^{2}=0.25$ for $k=1,2,\ldots,K$.

       In terms of entrance probabilities $\beta_{t-1}$, we consider the model
       \begin{equation*}
           \log \left(\frac{\beta_{t-1}}{\beta_{T-1}}\right)=\mbv{\Lambda}_{t}^{\prime}\mbv{\gamma},
       \end{equation*}
where $\mbv{\Lambda}_{t}=(\Lambda_{1t},\Lambda_{2t},\Lambda_{3t})^{\prime}$ for $t=1,2,\ldots,T-1$; and $\Lambda_{1}$, $\Lambda_{2}$, and $\Lambda_{3}$ are three time dependent covariates. These three covariates are simulated according to $\Lambda_{1t}, \Lambda_{2t}, \Lambda_{3t}\stackrel{iid}\sim \mbox{N}(0,1)$ for $t=1,2,\ldots,T$. For the regression coefficients $\mbv{\gamma}=(\gamma_{1},\gamma_{2},\gamma_{3})^{\prime}$, we consider $\mbv{\gamma}=(1.2,-0.8,0.6)^{\prime}$.

       In terms of the prior specification, we set $M=12,000$ and $a_{\Psi}=b_{\Psi}=1.0$. For regression coefficients $\mbv{\theta}$, $\mbv{\zeta}$, and $\mbv{\gamma}$, the prior distributions are given by: $\mbv{\theta}\sim \mbox{N}(\mbv{0},100\mbf{I}_{2})$, $\mbv{\zeta}\sim \mbox{N}(\mbv{0},100\mbf{I}_{3})$, and $\mbv{\gamma}\sim \mbox{N}(\mbv{0},100\mbf{I}_{3})$. For variance parameter $\sigma_{u}^{2}$, we consider $\sigma_{u}^{2}\sim \mbox{IG}(q_{u},r_{u})$ with $q_{u}=2.1$ and $r_{u}=1.1$. For parameters related to the OU process, we consider the prior specification as: $x_{0} \sim \mbox{N}(\mu_{0},\sigma_{0}^{2})$ where $\mu_{0}$ is the sample mean of observed values of $x_{it}$ based on captured individuals; $\sigma_{0}^{2}\sim \mbox{IG}(q_{\sigma_{0}},r_{\sigma_{0}})$ with $q_{\sigma_{0}}=2.1$ and $r_{\sigma_{0}}=1.1$; $\alpha \sim \mbox{N}(\mu_{\alpha},\sigma_{\alpha}^{2})$ with $\mu_{\alpha}=5$ and $\sigma_{\alpha}^{2}=100$; $\tau \sim \mbox{Unif}(q_{\tau},r_{\tau})$ with $q_{\tau}=.01$ and $r_{\tau}=5.0$; $\sigma^{2}\sim \mbox{IG}(q_{\sigma},r_{\sigma})$ with $q_{\sigma}=2.1$ and $r_{\sigma}=1.1$. Our prior specification reflects vague prior distributions relative to scale of the simulated data. 
              
       For the MCMC implementation, we run three chains in parallel each with a total of 150,000 iterations. For each Markov chain, we discard the first 100,000 iterations as burn-in and draw inference based on every fifth remaining samples. The convergence of the Markov chain to the stationary distribution is assessed via both trace plots of the sample chains and Gelman and Rubin's diagnostic \citep{brooks1998general}. In this case, visual inspection of the trace plots do not suggest lack of convergence for any model parameters. Moreover, the $\widehat{R}$ for all model parameters are less than 1.02.

\begin{figure}[ht]
   \centering
   \includegraphics[trim={5cm 1cm 5cm 0},clip,width=0.6\textwidth]{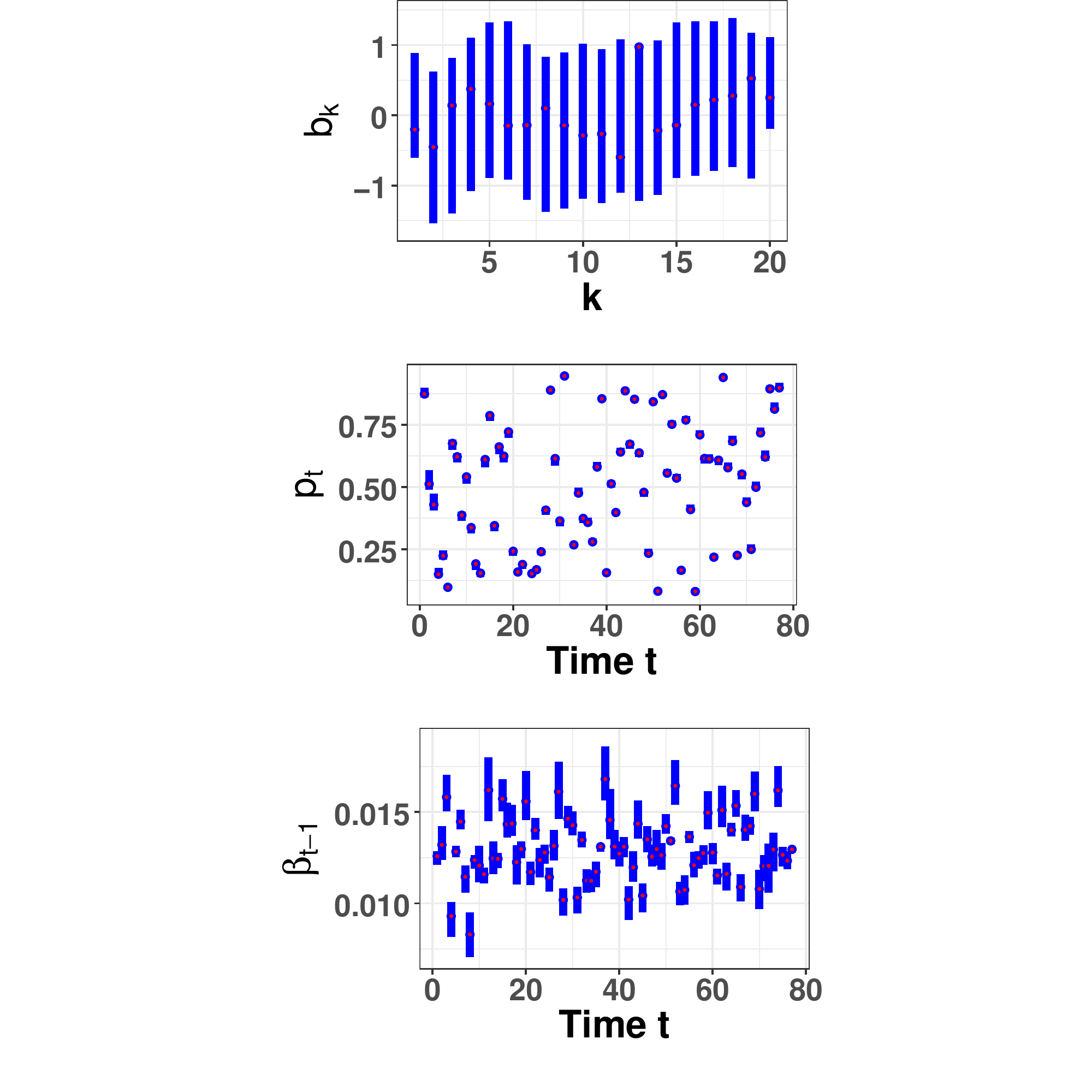}
   \caption{\baselineskip=10pt Plot of pointwise 95\% credible intervals and true values for $b_{k}$ ($k=1,2,\ldots,K=20$), capture probabilities $p_{t}$, and entrance probabilities $\beta_{t-1}$ for $t=1,2,\ldots,T$ in the simulated example (Section \ref{sec:Sim}). Note that the upper and lower value of each blue vertical line denotes the 2.5th and 97.5th percentiles of posterior samples, respectively. Also, the solid red circle on each blue line denotes the true value.} 
   \label{fig:JSInd_Simb}
\end{figure}

       Table~\ref{tab:JSInd_SimEst} provides posterior summary statistics for selected model parameters along with the corresponding true values. It can be seen that the 95\% credible intervals (CIs) cover the true values in all cases. In particular, for mean stopover duration $S$ and total stopover population size $N$, we can see from Table~\ref{tab:JSInd_SimEst} that their true values are recovered. For $b_{k}$, capture probabilities $p_{t}$, and entrance probabilities $\beta_{t-1}$, Figure~\ref{fig:JSInd_Simb} graphically presents their 95\% CIs along with the corresponding true values, from which we can conclude that all true values are recovered.
       
       For goodness-of-fit assessment, the Bayesian p-value for the JS model and the OU process is 0.35 and 0.46, respectively. Hence, these p-values do not suggest any lack-of-fit for either the JS model or the use of OU process. To summarize, this simulation suggests that we are able to correctly estimate parameters in the proposed model.

\section{Stopover Duration Analysis for Mallard}\label{sec:App}     
      We apply the model we propose to the stopover duration analysis for mallard, \textit{Anas platyrhynchos}. The mallard data was collected daily between August 1st and December 16th each year from 2004 to 2011. For illustration purpose, we only consider the mallard data collected in 2011. Moreover, we restrict our attention to the data collected between October 1st and December 16th since the number of daily captures prior to October is fairly low. For the data we consider, there are 686 individual mallards caught over $T=77$ days in 2011. Each day when a mallard was captured, measurements on body mass and structural size were taken, based on which body condition is calculated (as the ratio of body mass to structural size). Mallards have a determined growth, and once fully grown the structural size can be assumed to remain constant over time. 
               
      We consider the model for capture probabilities as
      \begin{equation}
         \mbox{logit}(p_{t})=\mbv{\Lambda}_{t}^{\prime}\mbv{\zeta}+\mbv{Z}_{t}\mbv{b},
         \label{eq:pcap_Mallard}
      \end{equation}
for $t=1,2,\ldots,T$. Here $\mbv{\Lambda}_{t}=(\mbox{PC}_{1t},\mbox{PC}_{2t},\mbox{PC}_{3t})^{\prime}$ and $\mbox{PC}_{1t}$, $\mbox{PC}_{2t}$, and $\mbox{PC}_{3t}$ are weather related covariates at day $t$ derived from the first three principal components (PCs) of the principal component analysis (PCA) conducted on five weather measures---wind direction, wind speed, atmospheric pressure, temperature, and rainfall. According to the PCA, three PCs explain about 80.4\% of the total variance. The first PC mainly reflects the dominant wind component (along the WSW-ENE axis) with negative values indicate strong WSW wind component whereas positive values indicate strong ENE wind component. The second PC reflects variation in rainfall with positive values indicate high atmospheric pressures associated with low precipitation. The third PC reflects the orthogonal wind component (along the NNW-SSE axis) and temperature deviation with positive values indicate strong NNW winds associated with temperatures below the seasonal norms, whereas negative values indicate strong SSE winds associated with temperatures higher than the seasonal norms. In addition, $\mbv{Z}_{t}$ is the $t$th row of matrix $\mbf{Z}=\mbf{Z}_{K}\mbf{\Omega}_{K}^{-\frac{1}{2}}$. Here $\mbv{Z}_{K}$ is the matrix with $t$th row $\mbv{Z}_{Kt}=(|t-\kappa_{1}|^{3}, |t-\kappa_{2}|^{3}, \ldots, |t-\kappa_{K}|^{3})^{\prime}$; and $\mbf{\Omega}_{K}$ is a $K\times K$ matrix whose $(k,l)$th entry is $|\kappa_{k}-\kappa_{l}|^{3}$ for $1\leq k,l\leq K$. Moreover, the $k$th fixed knot $\kappa_{k}$ is chosen as the sample quantile of $\{1,2,\ldots,T\}$ corresponding to probability $\frac{k}{K+1}$ for $k=1,2,\ldots,K$. The number of knots is chosen according to $K=\max\left\{20,\min\left(150,\frac{T}{4}\right)\right\}=20$.

      For departure and entrance probabilities, we consider two models as
      \begin{align*}
         \mbox{logit}(d_{i,t})&=\theta_{0}+\theta_{1}\texttt{Bcond}_{i,t}, \\
         \log\,\left(\frac{\beta_{t-1}}{\beta_{T-1}}\right)&=\mbv{\Lambda}_{t}^{\prime}\mbv{\gamma},
      \end{align*}
where $\texttt{Bcond}_{i,t}$ denotes the body condition for individual $i$ at day $t$ for $t=1,2,\ldots,T-1$. Since body condition varies with both individual and time, its change over time is modeled via the OU process discussed in Section \ref{sec:OU}.

      For the MCMC implementation, we set $M=2000$. In terms of prior distributions, we used the same specification as in Section \ref{sec:Sim}. We run three chains in parallel each with a total of 150,000 iterations. We discard the first 100,000 iterations as burn-in and summarize the posterior summary statistics based on every fifth remaining samples. To assess the convergence of the Markov chain to the stationary distribution, both trace plots of the sample chains and Gelman and Rubin's diagnostic are examined. In this case, visual inspection of the trace plots do not suggest lack of convergence for any model parameters. Moreover, the $\widehat{R}$ for all model parameters are less than 1.05.

      For the purpose of interpretation, we conclude that a parameter is significant if its 95\% CIs do not cover 0. Table~\ref{tab:JSInd_Mallard} provides posterior summary statistics for model parameters. According to Table~\ref{tab:JSInd_Mallard}, the posterior mean of coefficients $\zeta_{1}$ and $\zeta_{2}$ corresponding to the first two PCs are negative, which is opposite to that of the coefficient $\zeta_{3}$ for the third principal component. However, neither of these three coefficients are significant since their 95\% CIs all cover 0. For entrance probabilities, it is found that the third principal component has a significant negative effect on the timing of arrival for mallards because the 95\% CIs are entirely negative. This means that entrance probability increased when winds from SSE increases and temperatures exceeds the seasonal norms. As winds from SSE are opposite to tailwinds for mallard, mallard could choose to stop when winds do not provide flight assistance anymore, which would likely prevent them from skipping the stopover. For the total stopover population size, $N$, the result suggests that there were about 787 mallards that used our study area as a stopover site between October 1st and December 16th in 2011, with the corresponding 95\% CIs being [715,854]. For average stopover duration, $S$, the result suggests that, on average, mallards spent 11.4 days at our study site before flying to wintering areas, with the corresponding 95\% CIs of $S$ being [11.21,11.59].  
      
      Since the 95\% CIs of $\theta_{1}$ in Table~\ref{tab:JSInd_Mallard} are entirely negative, we conclude that there is a significant negative impact of body condition on departure probabilities. This result lends support to the necessity of incorporating body condition into the model for departure probabilities to account for individual heterogeneity in mallards' departure. In terms of impact of body condition on mallard departure decisions, our results suggest that birds have a high propensity to depart the stopover site when their body condition decreases. This result is somehow opposite to what is expected during stopover, whereby birds refuel fat stores (and increase body condition) in preparation for the next flight \citep{berthold2001bird}. One potential implication of such a finding could be that mallards experience poor refueling opportunities at our stopover site, e.g., due to insufficient food supply, or competition, forcing them to leave the site soon in searh for better refueling places \citep[e.g., see][]{schaub2008fuel}. Because changes in body condition are primarily due to changes in body mass, our result suggests that a mallard is more likely to leave our study site when its body mass decreases. Our finding surrounding the departure behaviors of mallards corroborates similar findings for migratory birds; e.g., see \citet{kuenzi1991stopover} and \citet{yong1993relation}. 
        
\begin{figure}[ht]
   \centering
   \includegraphics[scale=.8]{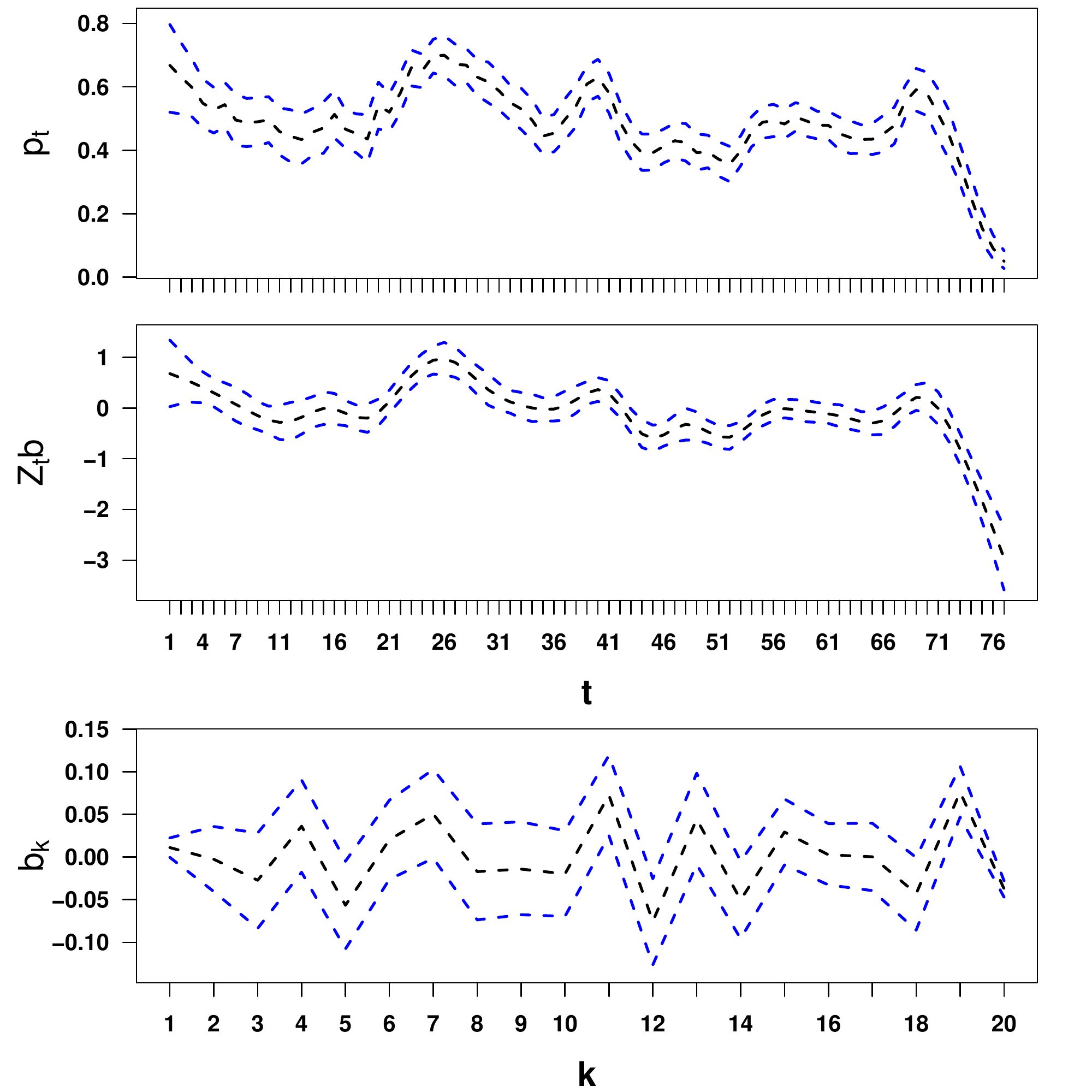}
   \caption{\baselineskip=10pt Plots of pointwise summary statistics of capture proabilities $p_{t}$, $\mbv{Z}_{t}b$, and $b_{k}$ for mallard (Section \ref{sec:App}). Note that the blue dashed lines are the pointwise 95\% credible intervals; the black dashed line is the posterior mean.}
   \label{fig:JSInda_Mallard}
\end{figure}

      Figure \ref{fig:JSInda_Mallard} provides pointwise posterior summary statistics for $p_{t}$, the nonparametric part $\mbv{Z}_{t}\mbv{b}$ in (\ref{eq:pcap_Mallard}), and $b_{k}$. According to Figure~\ref{fig:JSInda_Mallard}, we can conclude that the nonparametric part of our model for the capture probability (\ref{eq:pcap_Mallard}) is needed because the 95\% CIs for $b_{5}$, $b_{11}$, $b_{12}$, $b_{14}$, $b_{18}$, $b_{19}$, and $b_{20}$ exclude 0. Moreover, we note that both $p_{t}$ and $\mbv{Z}_{t}\mbv{b}$ demonstrate the same nonlinear trend with respect to time, indicating that capture probability is dominated by the the nonparametric part $\mbv{Z}_{t}\mbv{b}$. This is expected since three weather covariates in (\ref{eq:pcap_Mallard}) are found to be insignificant according to Table~\ref{tab:JSInd_Mallard}.

\begin{figure}[ht]
   \centering
   \includegraphics[scale=.8]{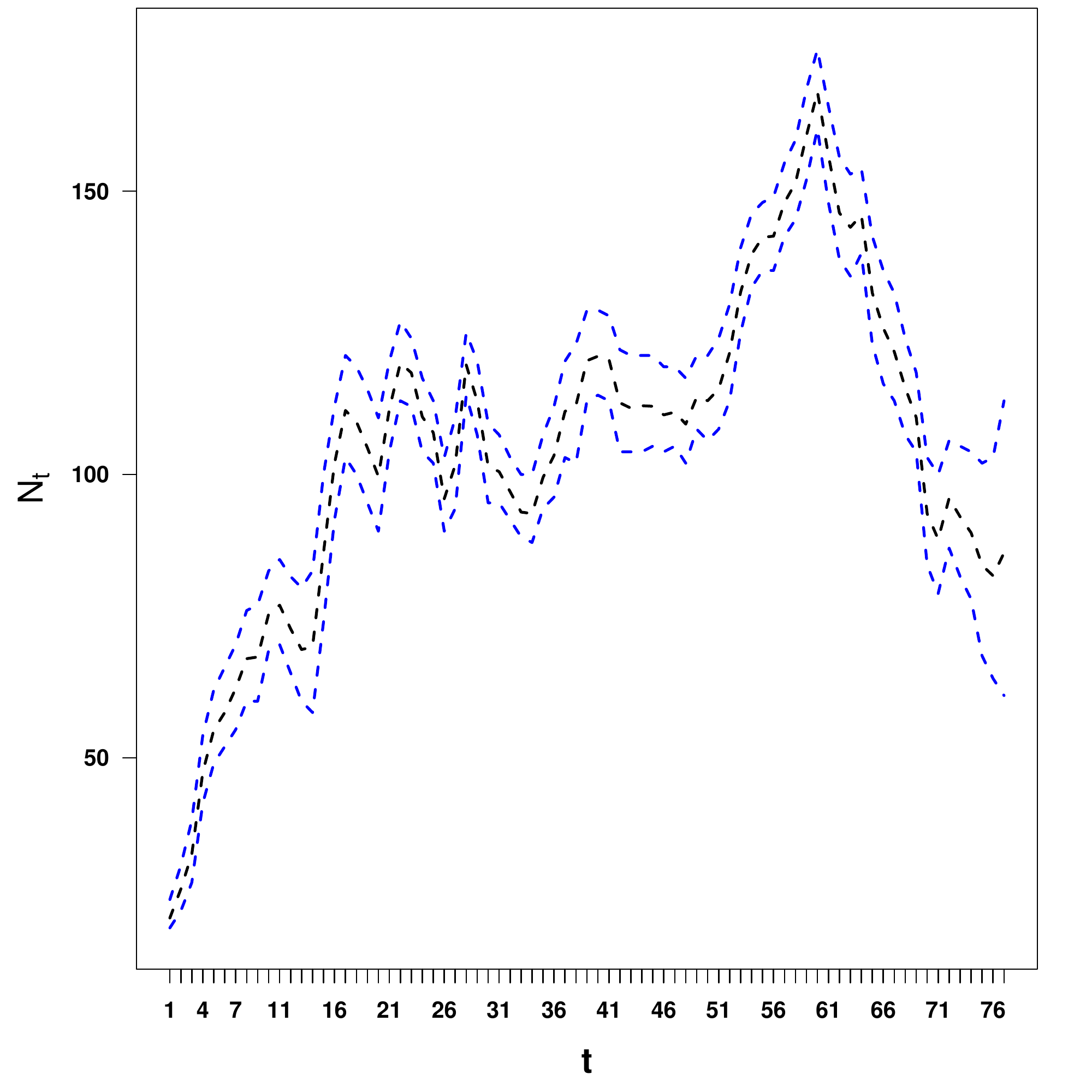}
   \caption{\baselineskip=10pt Plots of pointwise summary statistics of daily stopover population size $N_{t}$ for mallards (Section \ref{sec:App}). Note that the blue dashed lines are the pointwise 95\% credible intervals; the black dashed line is the posterior mean.}
   \label{fig:JSIndb_Mallard}
\end{figure}

      For the daily stopover population size, $N_{t}$, the corresponding pointwise posterior summary statistics are given in Figure~\ref{fig:JSIndb_Mallard}. From this figure, we can see that the daily stopover population sizes in October exhibit an overall upward trend. This upward trend in $N_{t}$ is repeated starting around mid-November to the very end of November, when daily stopover population sizes were peaked. Starting in December, there is an overall downward trend in $N_{t}$, suggesting that daily stopover population sizes decrease, which is due to the departure of mallards to wintering areas at this time of the season.

      Lastly, to assess goodness-of-fit of our proposed model, we compute the Bayesian p-values for both the JS model to the mallard data and for the modeling of body condition via the OU process. Using the complete log-likelihood function as the discrepancy function, the Bayesian p-value for the JS model is 0.39. In addition, the Bayesian p-value for the OU process is 0.61. Hence, these p-values do not suggest any lack of fit for either the JS model we propose for mallard data or the use of OU process.

\section{Discussion}\label{sec:Discu}
        Of particular importance to strategic management and conservation planning is to understand the contribution of various individual and environmental conditions to variation in stopover duration. In this paper, we propose a stopover model that is characterized by individual heterogeneity in departure, dependence of arrival time on covariates, and semiparametric modeling for the capture parameters based on the SA formulation of the JS model. To facilitate the design of the MCMC algorithm, the state-space formulation and data augmentation is adopted for our model. In the presence of a time-varying continuous individual covariate, the values of the covariate are partially observable at the times when each individual is captured. Thus, we have proposed the use of the OU process to model the change of such an individual covariate over time. 
        
        The model we propose has several distinct advantages. First and foremost, our model can be used to estimate the stopover duration, stopover population sizes, and to draw inference about how both intrinsic and extrinsic factors affect stopover departure behaviors for a specific species of interest, which are vital to many stopover duration analyses. Second, our model accounts for individual heterogeneity in departure due to a intrinsic factor that varies with both time and individual. Third, by linking entrance probabilities to extrinsic factors, we are able to examine the impacts of these factors on the time of arrival. Last but not least, we consider a semiparametric regression for capture parameters using low rank thin-plate splines, where the nonparametric part consists of a smooth function of time, allowing us to identify the functional relationship between time and capture probabilities. Collectively, these developments provide a framework with increased biological relevance that can be applied to any dataset with sufficient data. This has big premise for migration and movement ecology, as the length and behavior of birds during stopover are instrumental for overall speed of migration, with consequences at both the individual and the population level. As motivating example, we applied this model to capture-recapture data of mallards at an important stopover site in Sweden during fall migration. We were able to estimate stopover duration, stopover population size, the role of body condition on the timing of departure, as well as the impact of weather conditions on the timing of arrival.

        Despite the flexibility of semiparametric regression in our model, it is worth mentioning that its usage in capture-recapture analysis is not new. For example, \citet{gimenez2006semiparametric} consider Bayesian penalized splines that utilize truncated polynomial basis to model survival probabilities in the CJS model. Similarly, \citet{bonner2009time} consider a Bayesian semiparametric regression for survival probability in the CJS model that features B-spline basis functions. Instead of fixing the number and location of knot points, they consider a Bayesian adaptive approach where a reversible jump MCMC algorithm is employed to explore splines with different knot configurations. In addition, \citet{bonner2011smoothed} develop a model for data from Peterson-type mark-recapture experiment, where B-spline basis functions are considered to smooth population size estimates. In this paper, we apply Bayesian low rank thin-plate splines to smooth capture probabilities in the JS model. Unlike other basis functions, the use of low rank thin-plate splines leads to better mixing of the MCMC chains in Bayesian analysis \citep{crainiceanu2007bayesian}. Different from the Bayesian P-spline approach that often involves computational and numerical issues \citep[e.g., see][]{bonner2011smoothed}, we are able to develop a well-tailored sampling algorithm for our model that avoids any tuning through data augmentation. 
        
        To account for individual heterogeneity, \cite{bonner2009JSIndCov} develop a JS model that allows for individual heterogeneity in capture and survival probabilities due to a time-varying continuous individual covariate using a two-step Bayesian approach. The primary disadvantage of this two-step approach is that the entrance probability does not appear in the likelihood. Thus, one can not impose restrictions on/or link entrance probabilities to covariates, as in our proposed model. More critically, the two-step Bayesian approach relies on the careful specification of prior distributions in order to guarantee posterior distributions are well defined, which can impede its usage in practice. Similarly, \citet{schofield2011full} present a general framework for a variety of open population models with individual heterogeneity and demonstrate how freely available software programs, such as BUGS \citep{lunn2000winbugs}, can be used for Bayesian estimation of these models.  In principle, the authors provide a convenient platform for conducting many capture-recapture analyses; however, in practice, their approach has limited applicability. As acknowledged by the authors, their approach is limited to smaller datasets due to computational limitations; i.e., fitting large datasets using their approach can be inefficient. In contrast, we consider the data augmentation technique to facilitate the development of a customized sampling algorithm for model parameters. Specifically for latent variables, we propose block sampling algorithms that are extremely efficient. As a result, computationally, our methodology is applicable in the case of large datasets.

%\clearpage\pagebreak\newpage%\thispagestyle{empty}
\bibliographystyle{jasa}
\bibliography{Refinput}

\clearpage\pagebreak\newpage%\thispagestyle{empty}

\begin{table}
\centering
\begin{tabular}{crrrrrr}
  \hline
  Parameter               &$\mu_{\text{post}}$ & $\sigma_{\text{post}}$ & $Q_{.025}$ & $Q_{.50}$  & $Q_{.975}$ & truth           \\
  \hline
  $\zeta_{1}$ & 1.00276 & 0.01565 & 0.97198 & 1.00274 & 1.03407 & 1.0 \\ 
  $\zeta_{2}$ &-0.88542 & 0.01578 &-0.91650 &-0.88546 &-0.85450 &-0.9 \\ 
  $\zeta_{3}$ & 0.60098 & 0.01326 & 0.57515 & 0.60096 & 0.62683 & 0.6 \\ 
  $\theta_{0}$&-1.81683 & 0.01836 &-1.85323 &-1.81641 &-1.78132 &-1.8 \\ 
  $\theta_{1}$& 0.28110 & 0.02189 & 0.23841 & 0.28135 & 0.32324 & 0.3 \\ 
  $\gamma_{1}$& 1.29583 & 0.20374 & 0.90103 & 1.29550 & 1.69289 & 1.2 \\ 
  $\gamma_{2}$&-0.79020 & 0.21399 &-1.20903 &-0.78966 &-0.37420 &-0.8 \\ 
  $\gamma_{3}$& 0.61351 & 0.16912 & 0.28603 & 0.61225 & 0.94695 & 0.6 \\ 
  $x_{0}$     &-0.64975 & 0.13647 &-0.92522 &-0.65025 &-0.38476 &-0.64 \\ 
  $\sigma_{0}^{2}$ & 1.43562 & 0.16827 & 1.14986 & 1.41439 & 1.80354 & 1.37 \\ 
  $\alpha$    & 0.16691 & 0.04859 & 0.06829 & 0.16614 & 0.25531 & 0.2 \\ 
  $\tau$      & 0.18858 & 0.00674 & 0.17652 & 0.18807 & 0.20322 & 0.19 \\ 
  $\sigma^{2}$& 0.36079 & 0.01330 & 0.33536 & 0.36058 & 0.38729 & 0.36 \\ 
  $\sigma_{u}^{2}$  & 0.39473 & 0.23649 & 0.14643 & 0.33402 & 1.03127 & 0.25 \\ 
  \hline
\end{tabular}
\caption{\baselineskip=10pt Posterior summary statistics for parameters in the semiparametric JS model with individual heterogeneity for the simulated example presented in Section~\ref{sec:Sim}.}
\label{tab:JSInd_SimEst}
\end{table}

\begin{table}[htp]
\centering
\begin{tabular}{crrrrr}
  \hline
  Parameter   &$\mu_{\text{post}}$ & $\sigma_{\text{post}}$ & $Q_{.025}$ & $Q_{.50}$  & $Q_{.975}$ \\
  \hline
  $\zeta_{1}$  & -0.04927 & 0.05895 & -0.16485 & -0.04916 & 0.06408 \\ 
  $\zeta_{2}$  & -0.01396 & 0.03838 & -0.09001 & -0.01353 & 0.06075 \\ 
  $\zeta_{3}$  & 0.12381 & 0.06577 & -0.00196 & 0.12378 & 0.25329 \\ 
  $\theta_{0}$ &-1.24418 & 0.45653 &-2.19409 &-1.24998 &-0.36204 \\
  $\theta_{1}$ &-0.11037 & 0.04645 &-0.20006 &-0.11006 &-0.01485 \\ 
  $\gamma_{1}$ & -0.15395 & 0.08195 & -0.31704 & -0.15251 & 0.00327 \\ 
  $\gamma_{2}$ & 0.01892 & 0.06549 & -0.10454 & 0.01744 & 0.15349 \\ 
  $\gamma_{3}$ & -0.28984 & 0.08179 & -0.44985 & -0.29071 & -0.12737 \\ 
  $x_{0}$      & 10.69848 & 0.31305 & 10.12658 & 10.67671 & 11.34599 \\ 
  $\sigma_{0}^{2}$ & 0.95667 & 0.31854 & 0.49246 & 0.90416 & 1.74965 \\ 
  $\alpha$     & 9.78919 & 0.18521 & 9.42277 & 9.78049 & 10.14194 \\ 
  $\tau$       & 0.08946 & 0.00718 & 0.07525 & 0.08959 & 0.10349 \\ 
  $\sigma^{2}$ & 0.21371 & 0.01710 & 0.18699 & 0.21156 & 0.25223 \\ 
  $\sigma_{u}^{2}$   & 0.10073 & 0.03189 & 0.05595 & 0.09490 & 0.18088  \\  
  $N$          & 786.17830 & 36.07471 & 715 & 787 & 854 \\ 
  $S$          & 11.38919 & 0.09709 & 11.20700 & 11.38630 & 11.58601 \\   
 \hline
\end{tabular}
\caption{\baselineskip=10pt Posterior summary statistics for parameters in the semiparametric JS model with individual heterogeneity for the mallard data (Section \ref{sec:App}).}
\label{tab:JSInd_Mallard}
\end{table}   

\clearpage

\end{document}